\documentclass[11pt,tightenlines,nofootinbib]{revtex4}
\usepackage{amssymb}
\usepackage{amsmath}
\usepackage{amsfonts}
\usepackage[stable]{footmisc}
\usepackage{hyperref,blkcntrl,moredefs,relsize,attrib,tikz,booktabs,capt-of,float,varwidth,epsdice,graphicx}

\usetikzlibrary{decorations.pathmorphing,decorations.text}

\newenvironment{definition}[1][Definition]{\begin{trivlist}
\item[\hskip \labelsep {\bfseries #1}]}{\end{trivlist}}

\newcommand{\qed}{\nobreak \ifvmode \relax \else
      \ifdim\lastskip<1.5em \hskip-\lastskip
      \hskip1.5em plus0em minus0.5em \fi \nobreak
      \vrule height0.75em width0.5em depth0.25em\fi}

\let\baraccent=\= 
\renewcommand{\=}[1]{\stackrel{#1}{=}} 

\begin{document}

\title{Why is the universe comprehensible?}

\author{Ian T. Durham}
\email[]{idurham@anselm.edu}
\affiliation{Department of Physics, Saint Anselm College, Manchester, NH 03102}
\date{\today}

\begin{abstract}
Why is the universe comprehensible? How is it that we can come to know its regularities well-enough to exploit them for our own gain? In this essay I argue that the nature of our comprehension lies in the mutually agreed upon methodology we use to attain it and on the basic stability of the universe. But I also argue that the very act of comprehension itself places constraints on what we can comprehend by forcing us to establish a context for our knowledge. In this way the universe has managed to conspire to make itself objectively comprehensible to subjective observers.
\end{abstract}

\maketitle
\vspace{-12pt}
\section{Introduction}
Why is the universe comprehensible? How is it that we can come to know its regularities well-enough to exploit them for our own gain? This is one of the greatest enigmas of human knowledge. As Einstein once said, ``[t]he eternally incomprehensible thing about the world is its comprehensibility''~\cite{Einstein:1936aa}.\footnote{This translation of the phrase is attributed to Holton~\cite{Holton:1979aa}.} But, in order to comprehend something, Paul Davies has argued that the system being comprehended must necessarily have a certain level of organization that is embodied by the presence of non-random complexity within the system and that the act of comprehension requires accepting that this non-random complexity is in some sense objectively real~\cite{Davies:1990aa}. Objective reality is necessarily observer-independent. It is an inherently third-person perspective. Yet, if something is said to be comprehensible it is worth asking by whom or what; comprehensibility by its very definition implies the existence of something or someone doing the comprehending. Thus it would seem that comprehension is an inherently first-person, subjective act. Somehow the universe has conspired to make itself objectively comprehensible to subjective observers. 

Yet, these two perspectives are not necessarily incompatible. For example, Markus M\"{u}ller recently constructed a self-consistent theory in which an objective external world emerges from more fundamental observer states. The theory takes the first-person perspective as axiomatically true and, using the tools of algorithmic information theory, derives an emergent third-person perspective \textit{from} it~\cite{Muller:2019aa}. There is a sense in which a subjective first-person perspective is unavoidably fundamental. We are, after all, a part of this world that we seem to comprehend. We cannot step outside of it. In fact, our mere presence \textit{within} it is surely a regularity worthy of an explanation. But that still does not explain how it is that we can even countenance the possibility of arriving at an explanation in the first place. The very fact that the concept of explanation exists is, itself, in need of some explanation. 

To Einstein, `comprehensibility' meant a scientific understanding of the universe's functional composition. That is, he is presumably not interested in any metaphysical considerations. While metaphysical arguments for comprehensibility may exist, cf. theological explanations, these offer little in the way of reproducibility. It is arguably reproducibility that bridges the gap between the first-person, subjective act of comprehension and the third-person objective reality that is being comprehended. In a certain sense it could be said that objective reality is \textit{defined by} those regularities that can be independently reproduced or verified by different observers.

This question of the universe's comprehensibility, then, rests squarely on the ability of independent observers to reproduce or verify the existence of certain regularities. It is fundamentally an act of agreement, a willingness to acknowledge commonalities. In that sense, though comprehension is a subjective act for individual observers, for it to have any real meaning there must be some objective agreement by those observers about the regularities being comprehended. That is, the observers must, at least in a broad sense, share a common methodology for carrying out their observations. This makes the question of comprehensibility one of measurement. In this essay, I explore how the nature of this methodology itself helps to enable the comprehensibility of the universe while simultaneously placing limits on how much we can comprehend. 

\section{Comprehensibility}
The famous philosophical question\footnote{The origins of this question appear to be in George Berkeley's \textit{A Treatise Concerning the Principles of Human Knowledge} (1710). Its current form seems to have first been stated in~\cite{Mann:1910aa}.} ``if a tree falls in a forest and no one is there to hear it fall, does it make a sound?'' is usually interpreted as a metaphorical question about the nature of objective reality. Much less attention is paid to the nature of the question itself and what that implies for any hypothetical answer. A literal reading of the question and its sentence structure makes it clear that the question is about the sound produced (or not produced) by a falling tree. The tree is assumed to fall regardless of the presence or absence of a person in the vicinity. Even more fundamentally, there is never any question about the tree's existence. The tree exists and it has fallen. The question is whether the sound it makes in falling requires the presence of an observer.

Though the question concerning the tree is a truth-conditional sentence, not all questions necessarily take such a form. For example, the sentence ``what color is my hair?''\footnote{This is a trick question. I'm bald.} does not have a truth value. However, the \textit{answers} to questions are invariably declarative statements that \textit{do} have truth values. Thus questions can nearly always be understood within truth-conditional semantics~\cite{Hamblin:1958aa,Cross:2018aa}. But in order to judge the truth of a sentence we must understand its meaning~\cite{Tarski:1944aa}. According to Frege, there exist two primary types of expressions: proper names (which are nearly always singular terms) and functional expressions~\cite{Frege:1980aa}. Understanding the meaning of a statement requires understanding the context of these expressions. 

We can reformulate the question concerning the tree as a declarative statement to which a truth value may be assigned: ``any tree that falls in a forest will make a sound regardless of the presence or absence of an observer.'' The proper name in this sentence is `any tree that falls in a forest'. That is, this is a statement about a tree falling in a forest. It is \textit{not} a statement about a dog or the color yellow. The functional expression places the falling tree in a forest in context by associating it with the predicate `will make a sound regardless of the presence or absence of an observer'. Thus, this is a statement about whether a falling tree in a forest makes a sound in the presence or absence of an observer. It is \textit{not} a statement about whether a falling tree in a forest turns into a pigeon or reads a book. The ability to assign a truth value to the full statement rests on the fact that these two expressions each have referents, i.e. they specify objects, conditions, etc. Though the referents are necessary in the formulation of the statement, they naturally constrain the truth value. Simply put, we could ask an infinite number of questions about a tree falling in a forest, but as soon as we choose which question to ask, we have narrowed the context. This may seem trivial but it is actually deeply profound. While the question ``what color is my hair?'' could garner an infinite number of responses (e.g. `potato', `narrow', `blue', etc.) only a finite set of such responses makes logical sense. The statement ``any tree that falls in a forest will believe in Santa Claus'' is a nonsensical statement whose truth value has no real meaning.

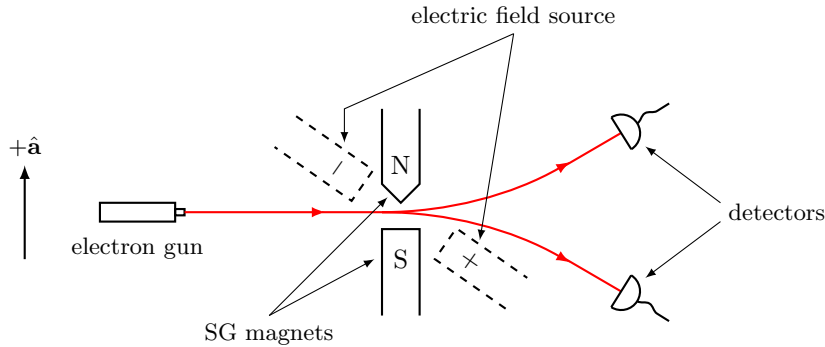
\begin{figure}
\begin{center}
\begin{tikzpicture}
\draw[thick,-latex] (-1,-0.5) -- (-1,0.75);
\node at (-1,1) {\footnotesize{$+\hat{\mathbf{a}}$}};
\draw[thick] (0,0) rectangle (1,0.25);
\draw[thick] (1,0.083) rectangle (1.125,0.167);
\node at (0.5,-0.35) {\footnotesize{electron gun}};
\draw[thick,red,-latex] (1.125,0.125) -- (3,0.125);
\draw[thick,red] (2.95,0.125) -- (4,0.125);
\draw[thick,red,-latex] (3.75,0.125) arc (90:60:5cm);
\draw[thick,red,-latex] (3.75,0.125) arc (270:300:5cm);
\draw[thick,red] (6.15,0.725) -- (7.05,1.25);
\draw[thick,red] (6.15,-0.475) -- (7.05,-1);
\node[rotate=32] at (7.25,1.375) {
	\begin{tikzpicture}
	\fill[white] (0,0) arc (-90:90:0.25cm);
	\draw[thick] (0,0) arc (-90:90:0.25cm);
	\draw[thick] (0,0) -- (0,0.5);
	\draw[thick,decorate,decoration={snake,amplitude=0.6mm,segment length=5mm}] (0.25,0.25) -- (0.75,0.25);
	\end{tikzpicture}
};
\node[rotate=-32] at (7.25,-1.125) {
	\begin{tikzpicture}
	\fill[white] (0,0) arc (-90:90:0.25cm);
	\draw[thick] (0,0) arc (-90:90:0.25cm);
	\draw[thick] (0,0) -- (0,0.5);
	\draw[thick,decorate,decoration={snake,amplitude=0.6mm,segment length=5mm}] (0.25,0.25) -- (0.75,0.25);
	\end{tikzpicture}
};
\node at (9,0.125) {\footnotesize{detectors}};
\draw[thin,-latex] (8.25,0.25) -- (7.25,1);
\draw[thin,-latex] (8.25,0) -- (7.25,-0.75);
\draw[thick] (3.75,-1.25) -- (3.75,-0.1) -- (4.25,-0.1) -- (4.25,-1.25);
\node at (4,-0.5) {S};
\draw[thick] (4.25,1.5) -- (4.25,0.5) -- (4,0.25) -- (3.75,0.5) -- (3.75,1.5);
\node at (4,0.75) {N};
\node at (2.25,-1.5) {\footnotesize{SG magnets}};
\draw[thin,-latex] (2.25,-1.25) -- (3.65,-0.5);
\draw[thin,-latex] (2.25,-1.25) -- (3.825,0.325);
\node[rotate=55] at (4,0.125) {
	\begin{tikzpicture}
	\draw[thick,dashed] (3.75,-1.75) -- (3.75,-0.6) -- (4.25,-0.6) -- (4.25,-1.75);
	\node at (4,-1) {$+$};
	\draw[thick,dashed] (4.25,2) -- (4.25,0.75) -- (3.75,0.75) -- (3.75,2);
	\node at (4,1.15) {$-$};
	\end{tikzpicture}
};
\node at (5.5,2.75) {\footnotesize{electric field source}};
\draw[thin,-latex] (5.5,2.5) -- (3.25,1.5) -- (3.25,1.05);
\draw[thin,-latex] (5.5,2.5) -- (5.05,-0.25);
\end{tikzpicture}
\caption{\label{spin} One method of measuring the spin of single electrons along a particular axis \textbf{a}, involves firing a properly attenuated beam of electrons through a Stern-Gerlach (SG) magnet and an appropriately applied transverse electric field. The transverse electric field is required to balance the Lorentz force. Each electron is then incident on one of two solid-state detectors.}
\end{center}
\end{figure}

In more concrete terms, consider an experiment designed to measure the spin of a single, free electron. One potential implementation of such an experiment consists of a source of single, free electrons (e.g. an electron gun), a Stern-Gerlach magnet, solid-state detectors, and a source of a transverse electric field to cancel out the transverse Lorentz force (see Figure~\ref{spin}). This measurement asks and answers the question ``what is the spin of an electron along the $\hat{\mathbf{a}}$ axis?'' We could restate the question as a truth-conditional statement of the form ``the electron is spin-up along the $\hat{\mathbf{a}}$ axis'' (alternately, we could state it as spin-down along the same axis). Physically speaking, the `proper name' (in Frege's terms) is the electron which is produced by the electron gun. All the other devices in the measurement generate the functional expression that places the electron in the context of a spin measurement along the $\hat{\mathbf{a}}$ axis. The choice of devices inherently determines the subject and predicate of the experiment. This experiment, for example, will not measure the spin of an electron along some axis $\hat{\mathbf{b}}$. In order to measure spin along an axis $\hat{\mathbf{b}}$, the magnetic and electric fields would need to be appropriately rotated. Neither will this experiment measure the spin of a silver atom along \textit{any} axis. This is not because the devices representing the functional expression can't accomplish the task, but rather it is because the source representing the proper noun (in Frege's terminology) only produces electrons. Even more obviously, this experiment will never determine the color of my hair without considerable modification.

Of course, one could simply deny these limitations and declare that the experiment \textit{has} actually determined the spin of an electron along some axis $\hat{\mathbf{b}}$ or the spin of a silver atom or the color of my hair. In other words, an individual observer could have an entirely different semantic interpretation of the experiment. This gets at the heart of science itself. As an anonymous reviewer in \textit{Philosophical Magazine} once declared (as quoted in~\cite{Eddington:1939aa}), ``[s]cience is the rational correlation of experience.'' That is, the \textit{objective} power of science lies in the correlation of multiple \textit{subjective} experiences. Each individual observer is free to declare whatever he or she wishes to declare. It is only when observers come to agree on elements of experience through rational means that objectivity is gained.\footnote{Due to length restrictions I will not endeavor to define `rational means' here. In relation to the aforementioned quote, Eddington expands on the idea in~\cite{Eddington:1939aa}.} Consider, for example, an experiment designed to test a Bell-type inequality using electron spin. We replace the electron gun in Figure~\ref{spin} with a source of entangled electron pairs that takes the form of two electron beams. Each beam passes through a setup similar to the one shown in Figure~\ref{spin} except the observer operating the equipment on a particular beam has the freedom to rotate the magnetic and electric fields to any desired direction thereby allowing for a measurement of spin along any axis in a plane orthogonal to the beam. By convention we refer to the two observers, each associated with one of the two beams, as Alice and Bob. Crucially, in order for this experiment to produce meaningful results, Alice and Bob must agree on the measurement protocol. That is, if their goal is to test a Bell-type inequality in such a way that they both agree to the result, then they must agree to the semantics of the statement whose truth value the experiment aims to determine. For instance, if Alice chooses to measure the spin of her electron along an axis $\hat{\mathbf{b}}$, she and Bob must have agreed to the definition of axis $\hat{\mathbf{b}}$ prior to carrying out the experiment if they have any hope of finding a meaningful result. But by doing so, they will have necessarily limited the context of the experiment since they will have defined axis $\hat{\mathbf{b}}$ to correspond to a specific direction. If they wish to achieve their stated goal, neither of them can simply deny the limitations of the setup. They \textit{must} agree on the semantic interpretation of the proper noun and functional expression in any truth-conditional statement they wish to jointly test \textit{since that is the point of the experiment}. This is another deeply profound point.

In a single-observer experiment, the observer is free to assign any meaning at all to the semantic content of the experimental statement. In the experiment described in Figure~\ref{spin}, for example, the observer could declare that the source produces silver atoms or even trees, for that matter, and simply assume that the results make meaningful sense. In other words, to a single observer, the truth value of the truth-conditional statement that the observer is testing does \textit{not} necessarily depend on that observer's interpretation of the statement. Observers in isolation are free to simply deny any result they obtain. Absent any independent verification, they could simply declare that the experiment produced a particular truth value. But in a two-observer experiment such as a Bell-type inequality test, the truth value of the statement being tested \textit{necessarily} depends on the mutually agreed upon interpretation of the statement they are testing. Suppose, for example, that Alice and Bob \textit{do} agree on the interpretation of the statement they are testing and suppose that statement is something to the effect of ``a series of $N$ runs of this experiment violates a Bell inequality.'' This means that they will agree on what constitutes a single run of the experiment, which Bell inequality is being tested (what a Bell inequality even \textit{is}, for that matter), what constitutes axis $\hat{\mathbf{a}}$, what constitutes axis $\hat{\mathbf{b}}$, etc. If, under these conditions, they act in good faith, then it means that they will agree on the truth value of the statement that they are testing. Note that if they have agreed to all these points (which constitute the functional expression of the statement) then (acting in good faith) they will necessarily agree on the result \textit{even if the equipment fails}. This is because they have agreed to the form of the functional expression itself. If you and I agree that a certain switch operates a light and we agree on what it means for a light to be illuminated when the switch is in a certain state, then we will agree (if acting in good faith) if the light fails to illuminate when the switch is in this state. We may disagree as to \textit{why} it failed to illuminate, but by agreeing to the conditions of the experiment we should agree on whether or not it \textit{has} illuminated. If we do not, then either one of us is not acting in good faith or there is some potentially hidden source of disagreement in the setup or execution.

Put another way, on a subsequent run of the experiment, Alice and Bob could fail to reach an agreement on what constitutes axis $\hat{\mathbf{b}}$. This disagreement could be known to them, but it also could be the result of experimental error. In either case, it is conceivable that Alice could find a violation of the inequality while Bob does not. This leaves the truth value of the statement that they are testing ambiguous. Each can \textit{say} that they individually comprehend the results of the experiment, but without some form of agreement, their individual interpretations remain nothing more than metaphysical speculation. That is, for their individual subjective comprehension to have any objective, scientific power, they must agree to what is being comprehended which means they must agree on the interpretation of the statement that they are testing. This is the only way in which the experiment can produce an unambiguous truth value. In other words, while individual observers can claim to observe anything, in order to determine unambiguous truth values of statements concerning the functional composition of the universe, observers must agree to the context of these statements and the protocol that is to be used to determine those truth values.

The comprehension that Einstein speaks of, then, arises from two interdependent conditions: 1) observers (of which there must be more than one) must agree on the meaning of truth-conditional statements and the methods used to test those statements, and 2) the results of the agreed upon tests must, in the aggregate, remain within the physical context of the tests themselves, i.e. an experiment such as the one depicted in Figure~\ref{spin}, for example, will not determine the color of my hair. In a certain sense, assuming a stable universe, the second condition follows from the first. If Alice and Bob agree to the meaning of a question they immediately limit the number of possible responses that will make any logical sense. If they then agree to the method of testing the question, they further limit the possible answers to be within the context of the test methodology. Put simply, if Alice and Bob agree to ask the question ``what color is my hair?'' and they agree on the methodology for testing that question, they can be fairly confident that the answer they find will not be ``narrow'' or ``dog''. As long as the universe remains relatively stable, the first condition implies the second.

The fact that the universe remains relatively stable, thus ensuring that the second condition is met in the aggregate, was originally proposed as a ``principle'' of comprehensibility in~\cite{Durham:2017aa}. It asserted that the nature of a physical system under investigation will always remain within the bounds of the method of investigation. That is, we expect scientific answers to scientific questions. Of course, one could object to the use of the word `always' as there is no way to prove this. In fact it is really a statement of tendencies within physical systems akin to the Second Law of Thermodynamics. Is it better to say that the entropy of the universe only \textit{tends} to increase rather than stating that it \textit{always} increases? I would rather hedge my bets and assume the weaker condition even if it has never been observed to decrease. Likewise with the principle of comprehensibility. In terms of truth-conditional statements, then, we can restate this principle of comprehensibility as follows.
\begin{definition}[Principle of Comprehensibility (Truth-conditional form)]
The results produced by tests of truth-conditional statements, for which the statements and the method of testing those statements are both agreed-upon by two or more observers, tend to remain within the physical context of the tests themselves.
\end{definition}
Again, despite being written here in terms of truth-conditional statements, which have a formal structure, this is a \textit{physical} principle. This is why we specify a `\textit{physical} context' within the principle's statement. It is not a formal (logico-mathematical) statement and thus has no rigorous proof. While it is impossible to \textit{prove} that the answer to the question ``what color is my hair?'' will never be ``narrow'' or ``dog'', the fact remains that it is highly unlikely. Even more unlikely is that ``narrow'' or ``dog'' will be an outcome of the experiment described in Figure~\ref{spin}. In short, stability and context lead to comprehension. Comprehension, however, comes with a price.

\section{The price of comprehensibility}
By simply asking a question, we immediately establish a context which limits the scope of the inquiry. The general stability of the universe ensures that answers to our question will, to a high degree of probability, remain within the context of the question. This is the essence of comprehensibility since it allows us to develop systematic measures by which we can further probe a topic. Building an understanding requires refining the context by asking additional questions. But this also necessarily means that our knowledge is shaped by the questions we ask and how we choose to ask them, i.e. by our methodology.

As a simple example, suppose we are tasked with designing a device that will compute a number by solving a specific equation. Choosing a design for the fundamental operation of our device is analogous to choosing a set of truth-conditional statements for a particular question. So, for example, perhaps we designed our device to be decimal (i.e. base-ten) at its most fundamental and it computes the number 1197. How this number is represented to us is immaterial. What matters is that the device was able to compute the number exactly based on how it was designed. We could have designed our device to be binary at its most fundamental, in which case it would compute the number 10010101101 since $10010101101_{2} = 1197_{10}$. Again, the representation of the number is immaterial here. Both devices could represent the number to us as base-eight or base-six or even as sounds or images. What matters is that the two designs are fundamentally different yet both ultimately produce the same solution to the equation with equal accuracy. The principle of comprehensibility ensures that the decimal device performs decimal operations while the binary device performs binary operations, i.e. the devices produce what they are \textit{designed} to produce. If they reliably and consistently continue to do so, we might infer that this pattern represents some kind of objectively real aspect of the world and we could say that we comprehended some element of it.

But suppose, instead, that our decimal device finds that the solution to the input equation is 1/10 and let us further suppose that this answer is exact. If our device was instead binary, the best it could do would be to approximate this number as a non-terminating expansion,
$$
\frac{1}{10} \approx \frac{1}{16}+\frac{1}{32}+\frac{1}{256}+\cdots =\frac{1}{2^4}+\frac{1}{2^5}+\frac{1}{2^8}+\cdots
$$
That is, the design of our device can potentially limit the accuracy of our knowledge and constrain our comprehension. If we are in doubt about the results we obtain, we could theoretically design a different device and check the results. If we began with our a binary device and realized the result was an approximation, we might be able to design a decimal device and compare the results of the two. If the decimal device naturally gives a fixed, terminating answer we can assume it is more accurate. But if we are, say, computing $\pi$, how would we know which is more accurate? 

This is not a trivial point. John Wrench and Levi Smith famously calculated the digits of $\pi$ using a gear-driven calculator by solving Machin's formula, eventually reaching 1120 digits by 1949~\cite{Richeson:2019aa}. Since then all improvements to their estimation have been on electronic computers. While decimal computers do exist and have been used to search for the digits of $\pi$ (e.g. the ENIAC), most of these machines were actually binary in their basic functionality and simply encoded decimal digits using some scheme such as binary-coded decimal (BCD) or excess-3. Since 2009 all records for the digits of $\pi$ have been carried out using Alexander Yee's multi-threaded y-cruncher program.\footnote{For details see http://www.numberworld.org/y-cruncher/.} Not only have we limited ourselves to electronic computers in our search for the digits of $\pi$ but we have recently limited ourselves to a single \textit{algorithm}. But as the simple example above demonstrates, the choice of \textit{device itself} can constrain the accuracy of the results. How do we know that a mechanical computer would necessarily give the same results? After all, a mechanical computer would not likely be able to execute the algorithm that is at the heart of Yee's y-cruncher program since that program is designed expressly for digital computers. But constructing a mechanical computer to calculate 50 trillion digits of $\pi$ (the current record) would likely be extremely difficult if not impossible. There are other reasons to trust Yee's algorithm and electronic computations in general, but the point is that there is really no way to prove just how accurate they are. Simply put, in order to compute a number (like $\pi$) we must build a device capable of carrying out the computation. But in doing so, we must choose a design and that choice immediately sets constraints on what types of results we can expect to get from the device. A binary device will produce binary results while a decimal device will produce decimal results.

The Church-Turing thesis conjectures that it ultimately should not matter what form the computation takes. The types of problems computable using one model should be computable using \textit{any} model since the thesis asserts that all computational models are equivalent to Turing machines. But the Church-Turing thesis is a formal conjecture. It's unclear if such a conjecture can necessarily be applied to physical systems~\cite{Copeland:2004aa,Copeland:2019aa} though several proposals for physical versions of the thesis have been made~\cite{Wolfram:2002aa,Piccinini:2011aa,Arrighi:2012aa}. But the physical world is a fickle beast. Proving something physically is very different from proving something mathematically. This point goes well beyond merely addressing Hilbert's Sixth Problem in which he called for an extension of the axiomatic methods of mathematics to physics~\cite{Hilbert:1902aa}. It is conceivable that, should a suitable theory of quantum gravity be found, we could eventually see a fully axiomatized physical \textit{theory}. But anyone who has spent any time in a laboratory will attest to the fact that the real world is far messier than theory would have us believe.

Comprehensibility is about what we, as humans, can know for certain about the universe and it is through experiment that we do this. As Richard Feynman once famously said, if something disagrees with experiment then it's wrong~\cite{Feynman:1965aa}. But the nature of comprehensibility ensures that the answers to the questions we pose to the universe are constrained by the form of the question itself. A \textit{physical} binary device, i.e. not an idealized abstract one, will always produce binary results and thus, for example, will never be able to calculate $1/10$ exactly. Certain \textit{physical} computations will always be approximate.

This has an even deeper and more profound implication if science ultimately rests entirely on proofs of truth-conditional statements. Proving a truth-conditional statement is equivalent to solving a decision problem or what Hilbert and Wilhelm Ackermann referred to as the \textit{Entscheidungsproblem}~\cite{Hilbert:1928aa}. Such a problem produces a yes-or-no answer to a question given any number of inputs. John Wheeler famously argued that the answers to such questions were the basis of all that exists~\cite{Wheeler:1990aa}. As late as 1930 Hilbert believed there would be never be an unsolvable problem~\cite{Hodges:1983aa}. Yet, though the Church-Turing thesis conjectures that all computational models are equivalent to Turing machines, it also allows for the existence of noncomputable functions. That is it allows for the existence of decision problems for which no algorithm can be constructed that is guaranteed to lead to a correct yes-or-no answer. Such problems are said to be undecidable. Turing, for instance, famously showed that the halting problem was undecidable on Turing machines~\cite{Turing:1936aa,Turing:1938aa}. But all of this work was grounded in formal methods. If the Principle of Comprehensibility is valid it would seem to imply that there might exist problems that are undecidable for \textit{physical} reasons. That is, if Wheeler is right and all that exists derives its very existence from ``apparatus-elicited'' answers to yes-or-no questions, then there are elements of the physical universe that are simply unknowable. Furthermore, the origin of that unknowability is both logical \textit{and} physical. Some aspects might be unknowable because we cannot construct an algorithm that is guaranteed to lead to a correct truth value for some truth-conditional statements. Other aspects might be unknowable because the universe's fundamental fabric is such that no machine can be constructed to produce a correct truth value for some truth-conditional statements. These are distinct points unless the universe itself is a Turing machine. The latter follows from the constraints associated with comprehensibility. In order to answer a question, we must choose a set of truth-conditional statements that we will test and agree to a testing methodology. This ensures that the results will be comprehensible. But, by choosing a particular set of such statements as opposed to some other set, we are shaping the form in which our answer will appear. When we ask a question we are necessarily carving out a small portion of knowledge from all that it is possible to know. Yet the question we ask and how we ask it immediately contextualizes any potential answer (even one that doesn't logically follow) and thus shapes any knowledge we gain from it. There is simply no way around this.

\section{Limitations}
The above arguments largely rest on the ability to reformulate the questions that lie at the heart of the scientific enterprise in terms of truth-conditional statements. But can all questions be reduced to truth-conditional statements? Can the scientific enterprise even be fully reduced to a set of questions? Attempts to formulate scientific explanation in terms of the answers to questions has a long history. Hempel and Oppenheim argued that all scientific explanations could be regarded as answers to `\textit{why}-questions' such as ``[w]hy do the planets move in elliptical orbits with the Sun at one focus?''~\cite{Hempel:1948aa}. Often referred to as the \textit{deductive-nomological} (DN) model, their account of the scientific enterprise is one of deduction that is reliant on the accurate prediction or postdiction of phenomena. But this model fell out of favor in the philosophy of science community for several decades and, despite a resurgence, continues to have its detractors chiefly because it appears to exclude some types of explanations generally regarded as scientific~\cite{Glennan:2006aa}. But even assuming that the scientific enterprise can be reduced to a set of questions, it is not clear that it can be further reduced to sets of truth-conditional statements. David Braun, for instance, offers an alternative in which the answers to questions simply provide contextual information~\cite{Braun:2006aa}. Truth is a loaded term. After all, how do we know that our knowledge of the world is even real? Would it not be better to simply speak of information? In Newton's day, testing the truth-conditional statement ``time is absolute'' would have produced a positive truth value given the knowledge and technology of the time. That same statement, if tested now, would elicit a negative truth value. Yet it is wrong to say that there was any change in the underlying physics between then and now. What changed was our \textit{knowledge} of that physics, i.e. we increased our information.

Yet even if we take Braun's view it is clear that limitations still exist and that these limitations intimately depend on context. Our inability to simultaneously measure non-commuting observables in quantum systems to arbitrary accuracy is a limit on our ability to obtain information regardless of whether or not we believe in objective truth. It arises from the context of our measurement. Comprehensibility is still the result of a combination of the mutual agreement between observers and the fact that the universe remains relatively stable. That is to say, the Principle of Comprehensibility need not be formulated in a truth-conditional form. The fact remains that in order to say we comprehend some element of the universe we must necessarily obtain some information about that element. But obtaining that information is a physical process that necessarily has a context which constrains the nature of that information; the very act of acquiring information shapes the information acquired. Physical limitations on the acquisition of knowledge are not controversial. The universe has fundamental limits baked into it. But it is these very limits that allow for the universe to be comprehensible. They are necessary in order for our seemingly finite minds to have any hope of comprehending anything.

These physical limitations bear a certain resemblance to G\"{o}del's incompleteness theorems in that they arise from the internal structure of the system and one must break free from that structure in order to fully understand it. The universe is a vast and interconnected place of which we are but a small part, a mere mote of dust, as it were. Any attempt to comprehend it must necessarily depend on the fact that we are a part of it. Indeed the very act of comprehension is \textit{itself} a part of it and is thus shaped by it. Like the god Odin from Norse mythology, who is said to have sacrificed an eye in order to attain wisdom, our quest for comprehension limits our very ability to comprehend, and the universe remains always partially veiled.

\newpage

\bibliographystyle{plain}
\bibliography{FQXi8.bib}

\end{document}